\documentclass[superscriptaddress,twocolumn,secnumarabic,
 amssymb,amsmath,nobibnotes,aps,prd,showkeys,showpacs,nofootinbib]{revtex4}

\usepackage{graphicx}
\usepackage{epsf}
\usepackage{bm}%
\newcommand{\be}{\begin{equation}}
\newcommand{\ee}{\end{equation}}

\newcommand{\mincir}{\raise
-3.truept\hbox{\rlap{\hbox{$\sim$}}\raise4.truept\hbox{$<$}\ }}
\newcommand{\magcir}{\raise
-3.truept\hbox{\rlap{\hbox{$\sim$}}\raise4.truept\hbox{$>$}\ }}

\begin{document}
\title{The Physical Properties of the Cosmic Acceleration}

\author{Spyros Basilakos}
\affiliation{Academy of Athens, Research Center for Astronomy and Applied Mathematics,
 Soranou Efesiou 4, 11527, Athens, Greece}

\begin{abstract}
The observed late-time acceleration of the cosmic expansion 
constitutes a fundamental problem in modern theoretical 
physics and cosmology. In an attempt to weight the validity of a
large number of dark energy models, I use the recent measurements 
of the expansion rate of the Universe, the clustering of galaxies
the CMB fluctuations as well as the large scale structure formation, 
to put tight constraints on the different models.
\end{abstract}
\maketitle

\section{Introduction}
Recent studies in observational cosmology lead to the conclusion that
the available high quality cosmological data
(Type Ia supernovae, CMB, etc.) are
well fitted by an emerging ``standard model''. This standard
model, assuming flatness, is described by the Friedman equation:
\be
H^2(a)=\left(\frac{{\dot a}}{a}\right)^2=\frac{8 \pi G}{3}\left[\rho_{m}(a)+
\rho_{X}(a)\right] \;,
\label{fe1}
\ee
where $a(t)$ is the
scale factor of the universe, $\rho_{m}(a)$ is the density
corresponding to the sum of baryonic and cold dark matter,
with the latter needed to explain clustering, and an extra component
$\rho_{X}(a)$ with negative pressure, called dark energy, needed to
explain the observed accelerated cosmic expansion
(eg., Davis et al. 2007; 
Kowalski et al. 2008; Komatsu et al. 2009 and references therein).

The nature of the dark energy is one of the most fundamental and
difficult problems in physics and cosmology. Indeed, during the
last decade there has been a theoretical debate among the cosmologists
regarding the nature of the
exotic ``dark energy''. Various candidates have been proposed in the
literature, such as a cosmological constant $\Lambda$ (vacuum),
time-varying $\Lambda(t)$ cosmologies, quintessence, $k-$essence,
vector fields, phantom, tachyons, Chaplygin gas and the list goes on
(for recent reviews see Peebles \& Ratra 2003;
Copeland, Sami \& Tsujikawa 2006; Caldwell \& Kamionkowski 2009). 
Within this framework, high energy field
theories generically indicate that the equation of state of such a
dark energy is a function of cosmic time. To identify this type
of evolution of the equation of state, a detailed form of the
observed $H(a)$ is required which may be obtained by a combination
of multiple dark energy observational probes.

In this cosmological framework, a serious issue 
is how (and when) the large scale structures form. 
Galaxies and large-scale structure grew 
gravitationally from tiny, nearly scale-invariant adiabatic 
Gaussian fluctuations. In this paper we focus also on galaxy
clusters that occupy an eminent position in the structure
hierarchy, being the most massive virialized systems known
and therefore they appear to be ideal tools for testing
theories of structure formation and extracting cosmological
information. The cluster distribution basically traces scales that have
not yet undergone the non-linear phase of gravitationally
clustering, thus simplifying their connections
to the initial conditions of cosmic structure formation.

The structure of the paper is as follows. In section 2 we 
briefly discuss the dark energy issue. In sections 3 and 4 we present 
the various dark energy models and
we use a joint statistical analysis, in order to
place constraints on the main cosmological parameters. In section
5 we present the corresponding theoretical
predictions regarding the formation of the galaxy clusters. 
Finally, we draw our conclusions in
section 6. Throughout the paper we will use
$H_{0}\simeq 71$km/sec/Mpc.

\section{The dark energy equation of state}
In the context of general relativity it is well known that
for homogeneous and isotropic flat cosmologies, driven by
non-relativistic matter and dark energy with equation
of state $p_X=w(a)\rho_X$, the expansion rate of the Universe
can be written as (see eq.\ref{fe1})
\begin{equation}
E^{2}(a)=\frac{H^{2}(a)}{H_{0}^{2}}=
\Omega_{m}a^{-3}+\Omega_{X}{\rm e}^{3\int^{1}_{a} d{\rm lny}[1+w(y)]} \;\;.
\label{nfe1}
\end{equation}
Note, that $E(a)$ is the normalized Hubble flow,
$\Omega_{m}$ is the dimensionless matter density
at the present epoch, $\Omega_{X}=1-\Omega_{m}$ is the
corresponding dark energy density parameter
and $w(a)$ is the dark energy equation
of state. Inverting the above equation we simply derive
\begin{equation}
\label{eos22}
w(a)=\frac{-1-\frac{2}{3}a\frac{{d\rm lnE}}{da}}
{1-\Omega_{m}a^{-3}E^{-2}(a)} \;\;.
\end{equation}
The exact nature of the dark energy has yet to be found and thus
the dark energy equation of state
parameter includes our ignorance regarding the
physical mechanism which leads to a late cosmic acceleration.

On the other hand, it is possible to extent the previous
methodology in the framework of modified gravity (see
Linder \& Jenkins 2003; Linder 2004). 
In this scenario, it is assumed that the dark energy may be 
an illusion, indicating the need to revise the general relativity and thus
also the Friedmann equation. From the mathematical point of view,
it can be shown that instead of using the exact Hubble
flow through a modification of the Friedmann equation we can
utilize a Hubble flow that looks like the nominal one (see
eq.\ref{fe1}). The key point here is to consider that the
accelerated expansion of the universe can be attributed to a
``geometrical'' dark energy component. Due to the fact that the
matter density in the universe (baryonic+dark) can not accelerate
the cosmic expansion, we perform the following parametrization
(Linder \& Jenkins 2003; Linder 2004):
\begin{equation}
E^{2}(a)=\frac{H^{2}(a)}{H_{0}^{2}}=
\Omega_{m}a^{-3}+\delta H^{2} \;\;.
\label{nfe2}
\end{equation}
Obviously, with the aid of the latter approach we include any
modification to the Friedmann equation of general relativity in
the last term of eq.(\ref{nfe2}). Now from eqs.(\ref{eos22},
\ref{nfe2}) we can derive, after some algebra, the ``geometrical''
dark energy equation of state
\begin{equation}
\label{eos222}
w(a)=-1-\frac{1}{3}\;\frac{d{\rm ln}\delta H^{2}}{d{\rm ln}a} \;\;.
\end{equation}
From now on, for the modified
cosmological models we will use the above formulation.

\section{Likelihood Analysis}
We will use various cosmological observations in order to
constrain the dark energy models described in section 4.
First of all, we use the  Baryonic Acoustic Oscillations
(BAOs). BAOs are produced by pressure (acoustic) waves in the
photon-baryon plasma in the early universe, generated by dark
matter overdensities. Evidence of this excess was recently found
in the clustering properties of the luminous SDSS red-galaxies
(Eisenstein et al. 2005; Padmanabhan et al. 2007) 
and it can provide a ``standard ruler'' with which we
can constraint the dark energy models. In particular, we use the
following estimator:
$
A({\bf p})=\frac{\sqrt{\Omega_{m}}}{[z^{2}_{s}E(a_{s})]^{1/3}}
\left[\int_{a_{s}}^{1} \frac{da}{a^{2}E(a)}
\right]^{2/3}$,
measured from the SDSS data to be $A=0.469\pm 0.017$, where $z_{s}=0.35$
[or $a_{s}=(1+z_{s})^{-1}\simeq 0.75$].
Therefore, the corresponding $\chi^{2}_{\rm BAO}$ function is simply written
\begin{equation}
\chi^{2}_{\rm BAO}({\bf p})=\frac{[A({\bf p})-0.469]^{2}}{0.017^{2}}
\end{equation}
where ${\bf p}$ is a vector containing the cosmological
parameters that we want to fit.

On the other hand, a very accurate and deep 
geometrical probe of dark energy is the
angular scale of the sound horizon at the last scattering surface as
encoded in the location $l_1^{TT}$ of the first peak of the
Cosmic Microwave Background (CMB)
temperature perturbation spectrum. 
This probe is described by the  CMB shift parameter
(Bond, Efstathiou \& Tegmark, 1997; Nesseris \& Perivolaropoulos 2007)  
and is defined as
$
R=\sqrt{\Omega_{m}}\int_{a_{ls}}^1 \frac{da}{a^2
E(a)}$.
The shift parameter measured from the WMAP 5-years
data (Komatsu et al. 2009) is $R=1.71\pm 0.019$ at $z_{ls}=1090$
[or $a_{ls}=(1+z_{ls})^{-1}\simeq 9.17\times 10^{-4}$].
In this case, the $\chi^{2}_{\rm cmb}$ function is given
\begin{equation}
\chi^{2}_{\rm cmb}({\bf p})=\frac{[R({\bf p})-1.71]^{2}}{0.019^{2}}
\end{equation}

Finally, we use the Union08 sample of
307 supernovae of Kowalski et al. (2008). The 
corresponding $\chi^{2}_{\rm SNIa}$ function becomes:
\begin{equation}
\label{chi22}
\chi^{2}_{\rm SNIa}({\bf p})=\sum_{i=1}^{307} \left[ \frac{ {\cal \mu}^{\rm th}
(a_{i},{\bf p})-{\cal \mu}^{\rm obs}(a_{i}) }
{\sigma_{i}} \right]^{2} \;\;.
\end{equation}
where $a_{i}=(1+z_{i})^{-1}$ is the observed scale factor of
the Universe, $z_{i}$ is the observed redshift, ${\cal \mu}$ is the
distance modulus ${\cal \mu}=m-M=5{\rm log}d_{L}+25$
and $d_{L}(a,{\bf p})$ is the luminosity distance
$
d_{L}(a,{\bf p})=\frac{c}{a} \int_{a}^{1} \frac{{\rm d}y}{y^{2}H(y)}
$
where $c$ is the speed of light.
We can combine the above probes by using a joint likelihood analysis:
${\cal L}_{tot}({\bf p})=
{\cal L}_{\rm BAO} \times {\cal L}_{\rm cmb}\times {\cal L}_{\rm SNIa} $
or
$\chi^{2}_{tot}({\bf p})=\chi^{2}_{\rm BAO}+\chi^{2}_{\rm cmb}+\chi^{2}_{\rm SNIa}$
in order to put even further constraints on the parameter space used.
Note, that we define the likelihood estimator \footnote{Likelihoods
are normalized to their maximum values. Note, that the step of sampling is 0.01
and the errors of the fitted
parameters represent $2\sigma$ uncertainties. 
Note that the overall number of
data points used is $N_{tot}=309$ and the degrees of freedom: 
$dof= N_{tot}-n_{\rm fit}$, with $n_{\rm fit}$ the number of fitted
parameters, which vary for the different models.} 
as: ${\cal L}_{j}\propto {\rm exp}[-\chi^{2}_{j}/2]$.

\section{Constraints on the flat dark energy models}\label{sec.constr}
In this section, we consider a large family of flat dark energy
models and with the aid of the above cosmologically relevant 
observational data, we attempt
to put tight constraints on their free parameters. In the
following subsections, we briefly present these cosmological
models which trace differently the nature of the dark energy.

\subsection{Constant equation of state - QP model}
In this case the equation of state is constant (see for a review,
Peebles \& Ratra 2003; hereafter QP-models). 
Such dark energy models do not have much physical motivation. In
particular, a constant equation of state parameter requires a fine
tuning of the potential and kinetic energies of the real scalar
field. Despite the latter problem, these dark energy models 
have been used in the literature due to their simplicity. 
Notice that dark energy models with a canonical kinetic term 
have $-1\le w<-1/3$. On the other hand, models of phantom dark energy
($w<-1$) require exotic nature, such as a scalar field with a negative 
kinetic energy. 
Now using eq.(\ref{nfe1}) the normalized Hubble parameter becomes
\be
E^{2}(a)=\Omega_{m}a^{-3}+(1-\Omega_{m})a^{-3(1+w)} \;\;.
\ee
Comparing the QP-models with the observational data
(we sample $\Omega_{m} \in [0.1,1]$ and $w \in [-2,-0.4]$)
we find that the best fit values are $\Omega_{m}=0.28\pm 0.02$ and
$w=-1.02\pm 0.06$ with
$\chi_{tot}^{2}(\Omega_{m},w)/dof \simeq 309.2/307$
in very good agreement with the
5 years WMAP data Komatsu et al. (2009). Also
Davis et al. (2007) utilizing 
the Essence-SNIa+BAO+CMB and a Bayesian statistics found
$\Omega_{m}=0.27 \pm 0.04$, while
Kowalski et al. (2008) using the Union08-SNIa+BAO+CMB obtained
$\Omega_{m}\simeq 0.285^{+0.03}_{-0.03}$. Obviously, our results
coincide within $1\sigma$ errors. It is worth noting that
the concordance $\Lambda$-cosmology can be described by QP models with
$w$ strictly equal to -1. In this case we find: $\Omega_{m}=0.28\pm 0.02$
with $\chi_{tot}^{2}(\Omega_{m})/dof\simeq 309.3/308$.

\subsection{The Braneworld Gravity - BRG model}
In the context of a braneworld cosmology (hereafter BRG) the accelerated
expansion of the universe can be explained by a modification of gravity
in which gravity itself becomes weak at very large distances (close to the
Hubble scale) due to the fact that our four dimensional brane
survives into an extra dimensional manifold 
(Deffayet, Dvali, \& Cabadadze 2002).
The interesting point in this scenario is that
the corresponding functional form of the normalized Hubble flow,
eq. (\ref{nfe2}), is affected only by one
free parameter ($\Omega_{m}$). Notice, that
the quantity $\delta H^{2}$ is given by
\be
\delta H^{2}=2\Omega_{bw}+2\sqrt{\Omega_{bw}}
\sqrt{\Omega_{m}a^{-3}+\Omega_{bw}} \;,
\ee
where $\Omega_{bw}=(1-\Omega_{m})^{2}/4$.
The geometrical dark energy equation of state parameter
(see eq.\ref{eos222}) reduces to
\be
w(a)=-\frac{1}{1+\Omega_{m}(a)} \;,
\ee
where $\Omega_{m}(a)\equiv \Omega_{m}a^{-3}/E^{2}(a)$.
The joint likelihood analysis provides a best fit value of
$\Omega_{m}=0.24\pm 0.02$, but the fit is rather poor
$\chi_{tot}^{2}(\Omega_{m})/dof\simeq 369/308$.

\subsection{The parametric Dark Energy model - CPL model}
In this model we use the Chevalier-Polarski-Linder
(Chevallier \& Polarski 2001; Linder 2003, hereafter CPL)
parametrization. The dark energy equation of state parameter is
defined as a first order Taylor expansion around the present
epoch:
\begin{equation}
w(a)=w_{0}+w_{1}(1-a) \;. \label{cpldef}
\end{equation}
The normalized Hubble parameter is given by (see eq.\ref{nfe1}):
\begin{equation}
E^{2}(a)=\Omega_{m}a^{-3}+(1-\Omega_{m})
a^{-3(1+w_{0}+w_{1})}e^{3w_{1}(a-1)} \;,
\end{equation}
where $w_{0}$ and $w_{1}$ are constants.
We sample the unknown parameters as follows: $w_{0} \in [-2,-0.4]$
and $w_{1} \in [-2.6,2.6]$.
We find that for the prior of $\Omega_{m}=0.28$ the overall
likelihood function
peaks at $w_{0}=-1.1^{+0.22}_{-0.16}$
and $w_{1}=0.60^{+0.62}_{-1.54}$
while the corresponding $\chi_{tot}^{2}(w_{0},w_{1})/dof$ is 307.6/307.

\subsection{The low Ricci dark energy - LRDE model}
In this modified cosmological model, we use a simple
parametrization for the Ricci scalar which is based on a Taylor
expansion around the present time: ${\cal R}=r_{0}+r_{1}(1-a)$
[for more details see Linder 2004]. It is interesting to
point that at the early epochs the cosmic evolution tends
asymptotically to be matter dominated. In this framework, we have
\begin{equation}
E^{2}(a)=\left\{ \begin{array}{cc}
        a^{4(r_{0}+r_{1}-1)}{\rm e}^{4r_{1}(1-a)} & \;\;\;\;a\ge a_{t}  \\
       \Omega_{m}a^{-3} & \;\;\;\;a<a_{t}
       \end{array}
        \right.
\label{SS}
\end{equation}
where $a_{t}=1-(1-4r_{0})/4r_{1}$. The
matter density parameter at the present epoch, is related
with the above constants via
$\Omega_{m}=\left(\frac{ 4r_{0}-4r_{1}-1 } {4r_{1}}
\right)^{4r_{0}+4r_{1}-1}
{\rm e}^{1-4r_{0}}$.
The equation of state parameter that corresponds 
to the current geometrical dark energy model is given by
\be
w(a)=\frac{1-4{\cal R}}{3}\left[1-\Omega_{m}
{\rm e}^{-\int_{a}^{1}(1-4{\cal R})(dy/y)}\right]^{-1} \;\;.
\ee
Note, that we sample the unknown parameters as follows: $r_{0} \in [0.5,1.5]$
and $r_{1} \in [-2.4,-0.1]$ and here are the results:
$r_{0}=0.82 \pm 0.04$
and $r_{1}=-0.74^{+0.10}_{-0.08}$ ($\Omega_{m} \simeq 0.28$) with
$\chi_{tot}^{2}(r_{0},r_{1})/dof \simeq 309.8/307$.

\subsection{The high Ricci dark energy - HRDE model}
A different than the previously described 
geometrical method was defined by 
Linder \& Cahn (2007),
in which the Ricci scalar at high redshifts evolves as
\be
{\cal R}\simeq \frac{1}{4}\left[1-3w_{1} \frac{\delta H^{2}}
{H^{2}}\right]\;,
\ee
where $\delta H^{2}=E^{2}(a)-\Omega_{m}a^{-3}$.
In this geometrical pattern the normalized Hubble flow becomes:
\be
E^{2}(a)=\Omega_{m}a^{-3}
\left(1+\beta a^{-3w_{1}}\right)^
{-{\rm ln}\Omega_{m}/{\rm ln}(1+\beta)} \;,
\ee
where $\beta=\Omega_{m}^{-1}-1$.
Using the same sampling as in the QP-models, the joint likelihood
peaks at $\Omega_{m}=0.28\pm 0.03$
and $w_{1}=-1.02\pm 0.1$ with
$\chi_{tot}^{2}(\Omega_{m},w_{1})/dof\simeq 309.2/307$. To this end,
the effective equation of state parameter
is related to $E(a)$ according to eq.(\ref{eos22}).

\subsection{The tension of cosmological magnetic fields  - TCM model}
Recently, Contopoulos \& Basilakos (2007) proposed a novel
idea which is based on the following consideration (hereafter TCM): if the
cosmic magnetic field is generated in sources (such as
galaxy clusters) whose overall dimensions
remain unchanged during the expansion of the Universe,
the stretching of this field by the cosmic expansion
generates a tension (negative pressure) that could possibly explain a small
fraction of the dark energy ($\sim 5-10\%$). In this flat
cosmological scenario the normalized Hubble flow becomes:
\be
E^{2}(a)=\Omega_{m}a^{-3}+\Omega_{\Lambda}+\Omega_{B}a^{-3+n}\;,
\ee
where $\Omega_{B}$ is the density parameter for the
cosmic magnetic fields and $\Omega_{\Lambda}=1-\Omega_{m}-\Omega_{B}$.
The equation of state parameter which is related
to magnetic tension is (see eq.\ref{eos22})
\be
w(a)=-\frac{3\Omega_{\Lambda}+n\Omega_{B}a^{-3+n}}
{3(\Omega_{\Lambda}+\Omega_{B}a^{-3+n})}\;\;.
\ee
To this end, we sample $\Omega_{B} \in [0,0.3]$ and $n \in [0,10]$
and we find that for the prior of $\Omega_{m}=0.28$ the best fit
values are: $\Omega_{B}=0.10\pm 0.10$ and $n=3.60^{+4.5}_{-2.6}$
with $\chi_{tot}^{2}(\Omega_{B},n)/dof\simeq 308.9/307$.

\subsection{The Pseudo-Nambu Goldstone boson - PNGB model}
In this cosmological model the dark energy equation of state
parameter is expressed
with the aid the potential $V(\phi)\propto [1+cos(\phi/F)]$ 
(Abrahamse et al 2008 and references therein):
\be
w(a)=-1+(1+w_{0})a^{F}\;,
\ee
where $w_{0} \in [-2,-0.4]$ and $F \in [0,8]$.
In case of $a\ll 1$ we get $w(a) \longrightarrow -1$.
Based on this parametrization the normalized Hubble function
is given by
\be
E^{2}(a)=\Omega_{m}a^{-3}+(1-\Omega_{m})\rho_{X}(a) \;.
\ee
In this context, the corresponding dark energy density is
\be
\rho_{X}(a)={\rm exp}\left[\frac{3(1+w_{0})}{F}(1-a^{F}) \right] \;\;.
\ee
Notice, that the likelihood function peaks at $w_{0}=-1.04\pm 0.17$
and $F=5.9\pm 3.2$ with $\chi_{tot}^{2}(w_{0},F)/dof\simeq 317/307$.

\subsection{The early dark energy - EDE model}
Another cosmological scenario that we include in our paper is
the  early dark energy model (hereafter EDE).
The basic assumption here is that at early epochs the amount of
dark energy is not negligible 
(Doran, Stern \& Thommes 2006 and references therein).
In this model, the total dark energy component is given by:
\be
\Omega_{X}(a)=\frac{1-\Omega_{m}-\Omega_{e} (1-a^{-3w_{0}})}
{1-\Omega_{m}-\Omega_{m}a^{3w_{0}}}+\Omega_{e}(1-a^{-3w_{0}})\;,
\ee
where $\Omega_{e}$ is the early dark energy density and $w_{0}$
is the equation of state parameter at the present epoch. Notice, that
the EDE model was designed to simultaneously (a) mimic the effects of the
late dark energy and (b) provide a decelerated expansion of the universe at high redshifts.
The normalized Hubble parameter is written as:
\be
E^{2}(a)=\frac{\Omega_{m}a^{-3}}{1-\Omega_{X}(a)}\;,
\ee
while using eq.(\ref{eos22}), we can obtain the
equation of state parameter as a function of the scale factor.
From the joint likelihood analysis we find that
$\Omega_{e}=0.05\pm 0.04$ and $w_{0}=-1.14^{+0.18}_{-0.10}$
(for the prior of $\Omega_{m}=0.28$) with
$\chi_{tot}^{2}(\Omega_{e},w_{0})/dof\simeq 308.7/307$.

\subsection{The Variable Chaplygin Gas - VCG model}
Let us consider now a completely different model namely the
variable Chaplygin gas (hereafter VCG) which corresponds
to a Born-Infeld tachyon action 
(Bento, Bertolami \& Sen 2004). Recently, an
interesting family of Chaplygin gas models was found to be consistent
with the current observational data 
(Dev, Alcaniz \& Jain 2003).
In the framework of a
spatially flat geometry,
it can be shown that the normalized Hubble function takes the following formula:
\begin{equation}
E^{2}(a)=\Omega_{b}a^{-3}+(1-\Omega_{b})
\sqrt{B_{s}a^{-6}+(1-B_{s})a^{-n}}\;,
\end{equation}
where $\Omega_{b}\simeq 0.021h^{-2}$ is the density
parameter for the baryonic matter 
(see Kirkman et al. 2003) and
$B_{s} \in [0.01,0.51]$ and $n\in [-4,4]$.
The effective matter density parameter is:
$\Omega^{eff}_{m}=\Omega_{b}+(1-\Omega_{b})\sqrt{B_{s}}$.
We find that the best fit parameters are
$B_{s}=0.05\pm 0.02$ and $n=1.58^{+0.35}_{-0.43}$ ($\Omega^{eff}_{m}\simeq 0.26$)
with $\chi_{tot}^{2}(B_{s},n)/dof \simeq 314.7/307$.

\section{Evolution of matter perturbations}
The evolution equation
of the growth factor for models where the dark energy
fluid has a vanishing anisotropic stress and the matter fluid is not
coupled to other matter species is given by
(Linder \& Jenkins 2003)
\begin{equation}
D^{''}+\frac{3}{2}\left[1-\frac{w(a)}{1+X(a)}\right]\frac{D^{'}}{a}-
\frac{3}{2}\frac{X(a)}{1+X(a)}\frac{D}{a^{2}}=0\;,
\label{deltatime1}
\end{equation}
where
\begin{equation}
X(a)=\frac{\Omega_{m}}{1-\Omega_{m}}{\rm e}^{-3\int_{a}^{1} w(y) d{\rm
    ln}y}=\frac{\Omega_{m}a^{-3}}{\delta H^{2}} \;.
\end{equation}
Note, that the prime denotes derivatives with respect
to the scale factor.
Useful expressions of the growth factor can be found for the
$\Lambda$CDM cosmology in Peebles (1993), for dark energy models with 
$w=const$ in Silveira, \& Waga (1994)
for dark energy models with
a time varying equation of state in Linder \& Cahn (2007)
and for the
scalar tensor models in Gannouji \& Polarski (2008).
In this work the growth factor
evolution for the current cosmological model is derived by solving
numerically eq. (\ref{deltatime1}).
Note, that the growth factors are normalized to unity
at the present time.

\subsection{The formation of galaxy clusters}
In this section we briefly investigate the cluster formation
processes by generalizing the basic equations
which govern the behavior of the matter perturbations
within the framework of the current dark energy models.
Also we compare our predictions with those found for the traditional
$\Lambda$ cosmology. This can help us
understand better the theoretical expectations of
the dark energy models as well as the variants from
the usual $\Lambda$ cosmology.

\begin{figure}[ht]
\mbox{\epsfxsize=10cm \epsffile{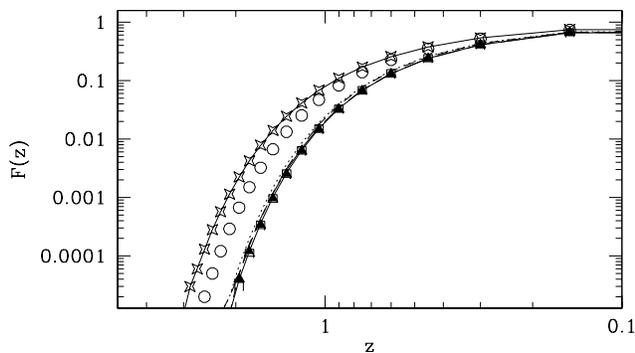}} \caption{The predicted fractional 
rate of cluster formation as a
function of redshift for the current cosmological models
($\sigma_{8}=0.80$). The points represent the following cosmological models: (a)
BRG (open stars), (b)
LRDE (open squares), (c)
TCM (open triangles), (d)
EDE (open circles) and (e)
PNGB (solid triangles).
The lines represent: (a)
CPL model (long dashed), (b) HRDE model (dot line)
and VCG model (dashed line).}
\end{figure}
The concept of estimating the fractional rate of cluster formation
has been proposed by different
authors (eg., Weinberg 1987; Richstone, Loeb \& Turner 1992). 
In particular, these
authors introduced a methodology which computes the
rate at which mass joins virialized structures, which grow from small
initial perturbations in the universe.
The basic tool is the Press \& Schechter (1974) 
formalism 
which considers the fraction of mass in the universe contained
in gravitationally bound structures (such as galaxy clusters)
with matter fluctuations greater than a critical value $\delta_{c}$,
which is the linearly extrapolated density
threshold above which structures collapse (Eke, Cole \& Frenk 1996).
Assuming that the density contrast is normally distributed
with zero mean and variance $\sigma^{2}(M,z)$
we have:
\be\label{eq:88}
{\cal P}(\delta,z)=\frac{1}{\sqrt{2\pi}\sigma(M,z)}
{\rm exp}\left[-\frac{\delta^{2}}{2\sigma^{2}(M,z)} \right] \;\;.
\ee
In this kind of studies it is common to
parametrize the rms mass fluctuation amplitude at
8 $h^{-1}$Mpc which can be expressed as a function of redshift as
$\sigma(M,z)=\sigma_{8}(z)=D(z)\sigma_{8}$.
The current cosmological models are normalized by
the analysis of the WMAP 5 years data
$\sigma_{8}=0.80$ (Komatsu et al. 2009).
The integration of eq.(\ref{eq:88}) provides the fraction of the universe,
on some specific mass scale, that has already
collapsed producing cosmic structures (galaxy clusters)
at redshift $z$ and is given by Richstone et al. (1992):
\be
\label{eq:89}
P(z)=\int_{\delta_{c}}^{\infty} {\cal P}(\delta, z) d\delta=
\frac{1}{2} \left[1-{\rm erf}
\left( \frac{\delta_{c}}{\sqrt{2} \sigma_{8}(z)} \right) \right]\;.
\ee
Notice, that for the model of modified gravity (BRG) we use
$\delta_{c}\simeq 1.47$ (see Schafer \& Koyama 2008),
for the EDE model
we use $\delta_{c}\simeq 1.4$ (see 
Bartelmann, Doran \& Wetterich 2006).
For the rest of the dark energy models,
due to the fact that $w \simeq -1$ close to the present epoch,
we utilize a $\delta_{c}$ approximation which is given by
Weinberg \& Kamionkowski (2003 see their eq.18).

Obviously the above generic form of eq.(\ref{eq:89})
depends on the choice of the background cosmology.
The next step is to normalize the probability to give the number of clusters which
have already collapsed by the epoch $z$ (cumulative distribution), divided
by the number of clusters which have collapsed at the
present epoch ($z=0$), $F(z)=P(z)/P(0)$.
In figure 1 we present, in a logarithmic scale, the behavior of normalized
cluster formation rate as a function of redshift for the
various dark energy models. In general, prior to $z\sim 0.2$ the cluster
formation has terminated
due to the fact that the matter fluctuation field, $D(z)$, effectively freezes.
For the traditional
$\Lambda$ cosmology we find the known behavior in
which galaxy clusters appear to be formed at high redshifts $z\sim 2$ 
(Basilakos 2003; Basilakos, Sanchez \& Perivolaropoulos 2009 
and references therein). 
From figure 1 it becomes also clear, that the vast 
majority of the dark energy models
seem to have a cluster formation rate which is close to that predicted
by the usual $\Lambda$ cosmology (see solid line in figure 1). 
However, for the BRG and EDE cosmological scenarios we find that galaxy
clusters appear to form earlier ($z\sim 3$) than in
the CPL, LRDE, HRDE, TCM, PNGB and VCG dark energy models.

\section{Conclusions}
In this work we have studied analytically and numerically the overall dynamics 
of the universe for a large number of dark energy models beyond the
{\em concrdance} $\Lambda$ cosmology, by using
several parameterizations for the dark energy equation of state.
We performed a joint likelihood analysis, using the current
high-quality observational data (SNIa, CMB shift parameter
and BAOs), and we put tight constraints on the main cosmological parameters.
We also find that for the vast majority of the dark energy models,
the large scale structures (such as galaxy clusters)
start to form prior to $z\sim 1-2$. 

{\bf Acknowledgments.}
For this paper, I have benefited from discussions with L.Perivolaropoulos 
and M. Plionis. Therefore, I would like to thank both of them.


\end{document}